\documentclass[conference]{IEEEtran}
\IEEEoverridecommandlockouts

\usepackage{pgfplots}
\usepackage{pgfplotstable}
\usetikzlibrary{pgfplots.statistics}
\usetikzlibrary{patterns}
\usepgfplotslibrary{groupplots}
\usepgfplotslibrary{fillbetween}
\usepackage{float}
\usepackage{algorithm2e}
\usepackage{rotating}
\usepackage{cite}
\usepackage{subcaption}
\usepackage{amsmath,amssymb,amsfonts}
\usepackage{algorithmic}
\usepackage{graphicx}
\usepackage{textcomp}
\usepackage{xcolor}
\usepackage{soul}
\usepackage[most]{tcolorbox}

\graphicspath{ {./images/} }

\def\BibTeX{{\rm B\kern-.05em{\sc i\kern-.025em b}\kern-.08em
    T\kern-.1667em\lower.7ex\hbox{E}\kern-.125emX}}
\begin{document}

\title{Delta Sum Learning: an approach for fast and global convergence in Gossip Learning
\thanks{Research funded by Flanders Research Foundation Junior Postdoctoral Researcher grant number 1245725N, and the NATWORK Horizon Europe project.}
}

\author{\IEEEauthorblockN{Tom Goethals, Merlijn Sebrechts, Stijn De Schrijver, Filip De Turck and Bruno Volckaert}
\IEEEauthorblockA{
\textit{Ghent University - imec, IDLab}\\
Gent, Belgium \\
ORCID: 0000-0002-1332-2290, 0000-0002-4093-7338, N/A, 0000-0003-4824-1199, 0000-0003-0575-5894}
}


\maketitle

\begin{abstract}

Federated Learning is a popular approach for distributed learning due to its security and computational benefits. With the advent of powerful devices in the network edge, Gossip Learning further decentralizes Federated Learning by removing centralized integration and relying fully on peer to peer updates. However, the averaging methods generally used in both Federated and Gossip Learning are not ideal for model accuracy and global convergence. Additionally, there are few options to deploy Learning workloads in the edge as part of a larger application using a declarative approach such as Kubernetes manifests. 
This paper proposes Delta Sum Learning as a method to improve the basic aggregation operation in Gossip Learning, and implements it in a decentralized orchestration framework based on Open Application Model, which allows for dynamic node discovery and intent-driven deployment of multi-workload applications. Evaluation results show that Delta Sum performance is on par with alternative integration methods for 10 node topologies, but results in a 58\% lower global accuracy drop when scaling to 50 nodes. Overall, it shows strong global convergence and a logarithmic loss of accuracy with increasing topology size compared to a linear loss for alternatives under limited connectivity.

\end{abstract}
\begin{IEEEkeywords}
gossip learning, artificial intelligence, ai, edge learning, federated learning
\end{IEEEkeywords}

\newcommand*{\circled}[2][]{\tikz[baseline=(C.base)]{
    \node[inner sep=0pt] (C) {\vphantom{1g}#2};
    \node[draw, circle, inner sep=1pt, yshift=1pt] 
        at (C.center) {\vphantom{1g}};}}

\section{Introduction}
\label{ch:intro}

Federated Learning (FL) has since long improved the training of Artificial Intelligence (AI) models by enabling distributed training on huge datasets while integrating the results at a central location, resulting in reduced training times and practical use of larger datasets. This is achieved through any number of integration methods, most popularly averaging methods (e.g. Federated Averaging or FedAvg). With the rise of edge computing, there is a push to further decentralize AI training through Gossip Learning (GL), which eliminates the centralized aggregation and relies on each node to spread its updates to the rest of the cluster by proxy, thus gossip. GL may be useful for various reasons; data may be processed at the source to reduce privacy concerns, spare computational capacity from dedicated hardware may be leveraged, etc. 
However, GL has certain inefficiencies compared to FL, for example, updates must be sent to several nodes instead of a single centralized location. Furthermore, because the distribution of training data and received updates is likely asymmetrical, models are likely to diverge during training. As a result, GL requires the integration of the full model whereas some FL approaches only send changed parameters to reduce network traffic. To counteract this, GL nodes should send updates to as few others as possible, which in turn leads to slower or possibly no global convergence, especially if not all nodes in a gossip cluster are mutually known. Furthermore, standard averaging approaches may result in statistical anomalies (e.g. vanishing variance), which lead to slower training convergence, as well as lower accuracy. As such, there is a need for integration methods that provide strong global convergence, through minimal and local communication with other nodes.

In terms of framework support, ML workloads are strongly supported in cloud environments. For example, KubeFlow enables MLOps in Kubernetes clusters, automating workload deployment and management using Kubernetes manifests on up to hundreds of nodes. However, edge learning lacks the same support; solutions such as Azure IoT Edge and OpenEI generally only enable edge intelligence, and in limited cases edge learning, whereas most recent studies focus on edge learning specifically through FL, or are aimed at specialized use cases that only leverage learning workloads. As such, there is a clear need for a framework that enables GL in the edge as a part of larger applications, while allowing Kubernetes-like modeling of workloads.

This paper examines the fundamental integration operation used by FL and GL, proposing Delta Sum Learning to improve global convergence. Additionally, this method is integrated into a decentralized intent-based framework for edge workload deployment, including Gossip and ML services to support GL.

Concretely, the contributions of this paper are:

\begin{itemize}
\item Improving Gossip Learning through Delta Sum Learning to provide better and faster global convergence with minimal local communication.
\item Integrating the improvements into a prototype framework for online edge learning and distributed updates.
\item Evaluating the efficiency and resource impact of the improvements compared to other Gossip Learning approaches.

\end{itemize}

The rest of this article is organized as follows: Section~\ref{ch:related} presents existing research related to the topic, while Section~\ref{ch:dslearning} introduces a mathematical model to improve Gossip Learning, which is integrated into a practical framework in Section~\ref{ch:flocky}. Section~\ref{ch:evaluation} details the evaluation setup and scenarios, while the results are presented in Section~\ref{ch:results}. Section~\ref{ch:discussion} presents topics for future work, and finally, Section~\ref{ch:finito} draws high level conclusions from the paper.

\section{Related Work}
\label{ch:related}

Federated Learning (FL) and by extension Decentralized Federated Learning (DFL) have been extensively studied\cite{Yuan2024} for their potential in Neural Network training. Specifically, a variety of model averaging or aggregation methods can be used, such as Federated Averaging, various Secure Aggregation\cite{Fereidooni2021} methods which mitigate privacy risks by obfuscating individual model updates, or Matched Averaging\cite{Shukla2021} methods aiming to reduce communication overhead. Furthermore, the effect of network topologies\cite{wang2019impact} and data distribution\cite{zhu2021federated} on the performance of (D)FL has been extensively studied, as well as mitigating noisy communications channels\cite{Chellapandi2024}.

Furthermore, a number of methods have been proposed that improve specific security aspects of (D)FL, including Byzantine-Robust FL\cite{Fang2024} for adversarial node behavior, or Distributionally Robust averaging\cite{Deng2020} which considers Non-IID data properties. Additionally, some methods aim to improve specific integration aspects, such as vanishing variance issues\cite{tian2024vanishing}. Finally, Xiao et al. argue that straightforward averaging methods are far from ideal for distributed learning\cite{Xiao2020}, showing an increasing correlation between nodes while computed distances between their models stop diminishing.

Gossip Learning\cite{9006216} (GL) has been proposed as a fully decentralized variant of DFL, using gossip algorithms to spread model updates throughout complex topologies. In GL, each node only directly contacts a small set of neighbours to push updates, usually through a gossip protocol such as SWIM\cite{1028914}. Gossip protocols may also benefit from push-pull operations\cite{10890446}, which is shown to offer superior update dissemination compared to push only as in most DFL approaches. Inheriting many (dis)advantages from DFL, its performance is shown to be on par with FL depending on hyperparameters, network data compression and integration methods\cite{HEGEDUS2021109}.

Network topology can have a significant impact on both (D)FL and GL; apart from the topologies discussed in \cite{wang2019impact}, more efficient update schemes can be constructed for FL when nodes are grouped in different silos\cite{NEURIPS2020_e29b722e}, and the decentralized nature of GL is shown to be robust in dynamic network environments\cite{di2023upsides}.

Existing frameworks such as EdgeFL\cite{ZHANG2025107600} offer a variety of options to run FL algorithms in the network edge, however, unlike the framework proposed in this paper they do not present a generic orchestration framework which allows ML workloads and GL to be used as components of a larger application.
There are a number of frameworks that solve more specific FL issues. For example, DeFTA considers the network outdegree of contributing nodes in DFL to weigh their updates\cite{ZHOU2024120582}, and integrates it into a DFL framework. Fedstellar\cite{MARTINEZBELTRAN2024122861} offers a similar framework, with various options to run DFL workloads, showing its effectiveness in attack detection.

\section{Delta Sum Learning}
\label{ch:dslearning}

This section builds a model for the improvement of the weight update integration step in Gossip Learning (GL), starting from the fundamental mechanism of batch updates on a single and multiple nodes. 


A single batch of training on one node can be represented as:

\begin{equation}
\label{eq:batch}
    w_{t+1} = w_{t} + \alpha \sum^{S}_{i=0} \frac{\partial{F(x_i,w_t,y_i)}}{\partial w_t}
\end{equation}

Where $w_t$ are the model weights at time $t$, $\alpha$ is the learning rate, $S$ iterates over the samples in a batch, and $F(x_i,w_t,y_i)$ is the model loss function using sample input $x_i$ and expected output (e.g. classification) $y_i$. Explaining this intuitively, the partial derivative is the slope vector at the hypersurface point defined by the model loss function $F(...)$ as an outcome of the model weights $w_t$, with respect to the weights $w_t$, which is exactly the factor required for gradient descent. Although multiple model layers are not explicitly represented here, $F(...)$ can represent any and all layers.

Extending this to multiple training epochs $T$, as well as a group of $N$ nodes simultaneously training on parts of the same dataset (i.e. FL), the expected integration becomes:

\begin{equation}
\label{eq:multiepoch}
    w_{t_0+T} = w_{t_0} + \alpha \sum^{t_0+T}_{t=t_0} \sum^{N}_{n=0} \sum^{S}_{i=0} \frac{\partial{F(x_{n,i},w_{n,t},y_{n,i})}}{\partial w_t}
\end{equation}

While there is a discrepancy here in the slightly different weights matrices $w_{n,t}$ learned by each node compared to $w_{t}$ in monolithic learning, for FL there is little difference between this and training in (large) batches, as the weight matrix of each node can be periodically synchronized with the centralized main model to keep $w_{n,t}$ from diverging. This is not the case with GL, however there is an issue with averaging methods to be considered first. In averaging approaches using full model updates, an update $u_{r,t}$ is received from each node every round, and integrated into a main model. Although FL often works by sending only weight tensor deltas instead of the entire model, the nature of GL requires the entire model be sent as there is no global model to integrate into; for now full model updates are assumed. The total received update $u_{a,t}$ is then:

\begin{equation}
    u_{a,t} = u_{l,t} + u_{r,t} 
\end{equation}

Where $u_{l,t}$ is the local learning update (if any), and $u_{r,t}$ is the combined update from remote nodes. These can be expressed in terms of Eq. \ref{eq:batch}, resulting in (respectively):

\begin{equation}
    u_{l,t} = w_{a,t} + \alpha \sum^{S}_{i=0} \frac{\partial{F(x_i,w_t,y_i)}}{\partial w_t}
\end{equation}

\begin{equation}
   u_{r,t} = \sum^{N}_{n=0} \left(w_{n,t} +  \alpha  \sum^{S}_{i=0} \frac{\partial{F(x_{n,i},w_{n,t},y_{n,i})}}{\partial w_{n,t}}\right)
\end{equation}

Or, when updates are the result of multiple learning epochs $T$, the analog of Eq. \ref{eq:multiepoch} becomes:

\begin{equation}
\label{eq:multinodeupdate}
\begin{split}
    u_{a,t_0 + T} =  & w_{a,t_0} + \alpha  \sum^{t_0+T}_{t=t_0} \sum^{S}_{i=0} \frac{\partial{F(x_i,w_{a,t},y_i)}}{\partial w_{a,t}} \\ 
 & +  \sum^{N}_{n=0}  \left(w_{n,t_0} + \alpha \sum^{t_0+T}_{t=t_0} \sum^{S}_{i=0}  \frac{\partial{F(x_{n,i},w_{n,t},y_{n,i})}}{\partial w_{n,t}}\right)
\end{split}
\end{equation}

Using this for an update of node 0 from epoch 0 to $T$, since all models $w_{n,t_0}$ are the same as $w_{a,t_0}$ at $t=0$, this reduces to:

\begin{equation}
\begin{split}
    u_{0,T} = w_{0,0}(N + 1) & + \alpha  \sum^{T}_{t=0} \sum^{S}_{i=0} \frac{\partial{F(x_i,w_{0,t},y_i)}}{\partial w_{0,t}} \\ 
 & +  \alpha  \sum^{N}_{n=0}  \sum^{T}_{t=0} \sum^{S}_{i=0}  \frac{\partial{F(x_{n,i},w_{0,t},y_{n,i})}}{\partial w_{0,t}}
\end{split}
\end{equation}

Or, substituting local and remote learning update factors with $\delta_{l,T}$ and $\delta_{r,T}$, respectively:

\begin{equation}
    u_{0,T} = w_{0,0}(N + 1)  +  (\delta_{l,T} + \delta_{r,T} )
\end{equation}

Which, when applying any averaging method results in a new model weight tensor $w_{0,T}$:

\begin{equation}
\label{eq:divupdate}
    w_{0,T} = w_{0,0}  + \frac{  (\delta_{l,T} + \delta_{r,T} )}{N+1}
\end{equation}

However, this shows that any straightforward averaging method imposes an implicit learning penalty based on the number of node updates received, as the actual updates are implicitly divided by $N+1$. Although this approach assumes updates with full model weight tensors, the result is analogous for FL with federated averaging using only model weight updates:

\begin{equation}
    w_{t+1} = w_{t} + \frac{ \sum^{N}_{n=0} k_n \Delta w_{n,t}}{\sum^{N}_{n=0} k_n}
\end{equation}

This method allows for different numbers of samples $k_n$ per node per update, while expressing the sum over multiple epochs $T$ from Eqs. \ref{eq:batch} and \ref{eq:multiepoch} as $\Delta w_{n,t}$ for readability. Expressed differently:

\begin{equation}
\label{eq:flresult}
    w_{t+1} = w_{t} + \frac{ \sum^{N}_{n=0} k_n \Delta w_{n,t}}{\mu_{kn} N}
\end{equation}

Where $\mu_{kn}$ is the average number of samples per node. Eq. \ref{eq:multiepoch} implicitly has the same number of learning samples for every batch and node; to allow for different numbers of learning samples per node it can be rewritten in similar terms as Eq. \ref{eq:flresult}:

\begin{equation}
    w_{t+1} = w_{t} + \frac{ \sum^{N}_{n=0} k_n \Delta w_{n,t}}{\mu_{kn}}
\end{equation}

Or, extending $\Delta w_{n,t}$ back to its full form:

\begin{equation}
    w_{t_0+T} = w_{t_0} + \frac{\sum^{N}_{n=0} k_n \sum^{t_0+T}_{t=t_0} \sum^{S}_{i=0} \alpha \frac{\partial{F(x_{n,i},w_{n,t},y_{n,i})}}{\partial w_t}}{\mu_{kn}}
\end{equation}

This approach still weighs node updates relative to the training samples for each update, while avoiding $N$ in the denominator.
However, due to differences in local model weights during training in GL, full model updates must be considered for further progress. To eliminate the $N+1$ factor from the updates themselves in Eq. \ref{eq:divupdate} and achieve a similar outcome to Eq. \ref{eq:multiepoch}, the base weight tensor and update tensor for each node must be separately transmitted during training updates, allowing for:

\begin{equation}
    w_{0,T} = w_{0,0}  +  (\delta_{l,T} + \delta_{r,T} )
\end{equation}

Or, expanding the $\delta$ factors again and expressed in a more general form:

\begin{equation}
\begin{split}
    w_{a,t_0 + T} = & \frac{w_{a,t_0} + \sum^{N}_{n=0} w_{n,t_0}}{N+1}  + \alpha  \sum^{t_0+T}_{t=t_0} \sum^{S}_{i=0} \frac{\partial{F(x_i,w_{a,t},y_i)}}{\partial w_{a,t}} \\ 
 & + \alpha  \sum^{N}_{n=0}   \sum^{t_0+T}_{t=t_0} \sum^{S}_{i=0}  \frac{\partial{F(x_{n,i},w_{n,t},y_{n,i})}}{\partial w_{n,t}}
\end{split}
\end{equation}

As the interiors of both the local (double) and remote (triple) sums represent a number of epochs on the same nodes n, they can be summarized as $w_{n,t_0+T}-w_{n,t_0}$ (with $\alpha$ already integrated in the learning algorithm):

\begin{equation}
\begin{split}
    w_{a,t_0 + T} =  \frac{w_{a,t_0} + \sum^{N}_{n=0} w_{n,t_0}}{N+1}  & +  (w_{a,t_0+T} - w_{a,t_0}) \\
 & + \sum^{N}_{n=0} (w_{n,t_0+T} - w_{n,t_0}) \\ 
\end{split}
\end{equation}

Or, when interpreting the local weight matrix and delta as node ``$N+1$'':

\begin{equation}
\label{eq:finally}
    w_{a,t_0 + T} =  \frac{ \sum^{N+1}_{n=0} w_{n,t_0}}{N+1} + \sum^{N+1}_{n=0}\Delta w_{n,t_0+T} 
\end{equation}

Eq. \ref{eq:multinodeupdate} shows that the models diverge after the first gossip update, especially since not all nodes are in direct contact. As a result, models will train into slightly different directions depending on their own data subset and initial updates received, especially during the first training rounds when accuracy and loss change rapidly when starting from randomized weight tensors. Therefore, all $w_{n,t}$ are likely to be minutely different throughout the learning process, although they should converge as updates spread through the gossip network, and they are assumed compatible due to training for the same global objective. Despite this, the derived model in  Eq. \ref{eq:finally} shows the necessity to integrate the base weight matrices from all nodes for which updates were received through averaging; Eq. \ref{eq:multiepoch} uses only the local matrix because all nodes are assumed to train on the same base matrix (periodically updated if necessary). 

\begin{figure*}[htbp]
  \centering
  \includegraphics[width=0.8\textwidth]{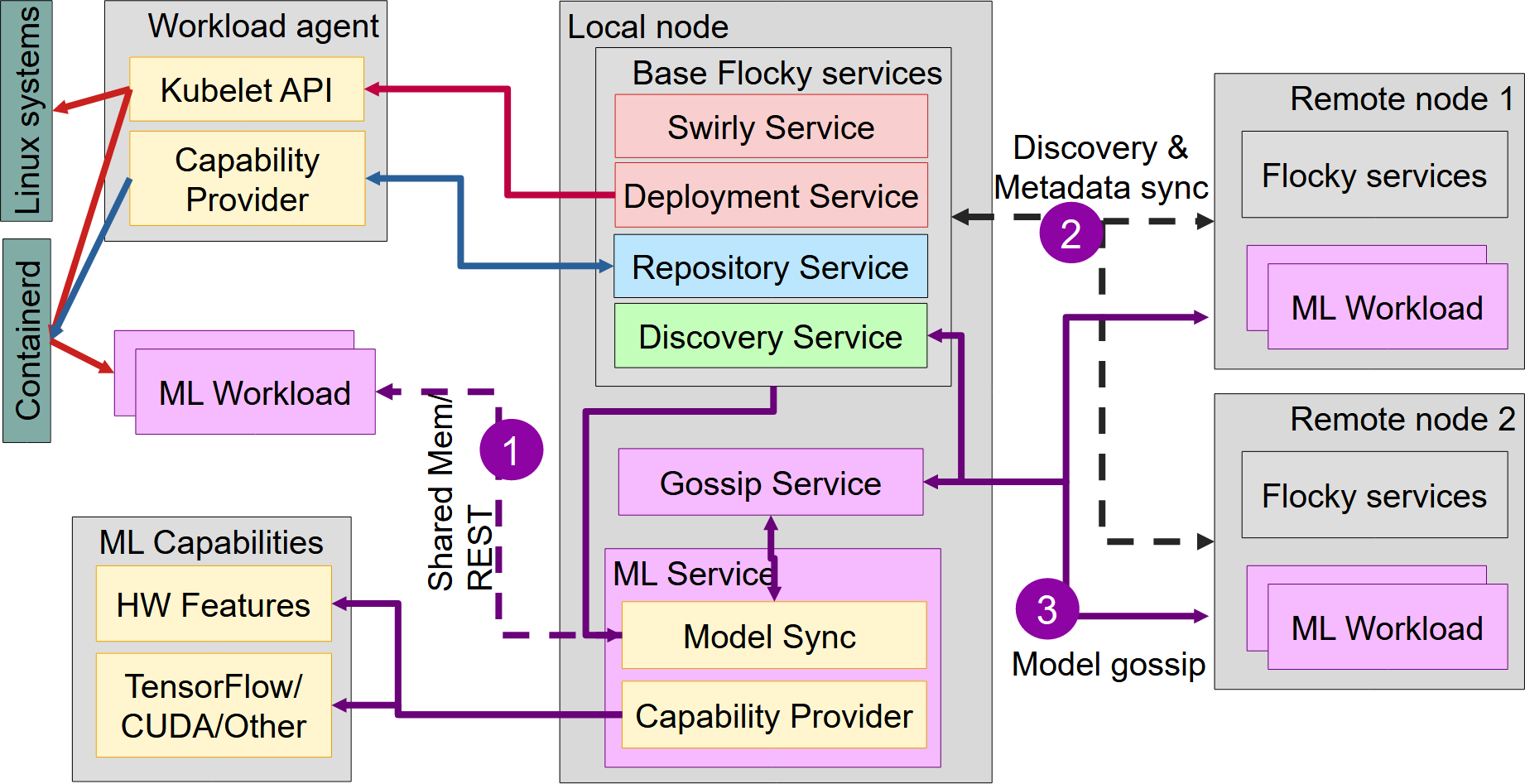}
  \caption{ Architecture overview of decentralized cluster discovery and Gossip Learning showing: \circled{1} Node discovery and metadata synchronization between nodes. \circled{2} ML model synchronization between workloads and a dedicated service. \circled{3} Model gossiping based on locally discovered nodes and hosted ML workloads. }
  \label{fig:arch}
\end{figure*}

Finally, there are some practical concerns. Because data and learning are not ideally spread in GL, there may be significant accuracy and loss jitter when receiving updates from nodes with contradictory updates. To reduce these effects, it is prudent to introduce a gossip learning factor $\lambda$:

\begin{equation}
    w_{a,t_0 + T} =  \frac{ \sum^{N+1}_{n=0} w_{n,t_0}}{N+1} + \lambda \sum^{N+1}_{n=0}\Delta w_{n,t_0+T} 
\end{equation}

This factor should be small enough to dampen erratic updates, yet large enough that its effect does not fall below $1/(N+1)$, which would slow down learning more than standard averaging methods already do. An acceptable value is determined in the evaluation section.
However, as learning topologies grow, nodes possess a relatively decreasing subset of the total learning set to train on; if training from local data does not perfectly align with the global optimum then training will have an increasingly detrimental effect on gossip updates, potentially canceling them out as:

\begin{equation}
\begin{split}
w_{a,t_0} + \Delta w_{a,t_0+T}  =\frac{ \sum^{N}_{n=0} w_{n,t_0}}{N}   + \sum^{N}_{n=0} (\Delta w_{n,t_0+T} )
\end{split}
\end{equation}

Which may be partially avoided using a dynamic learning factor; models are assumed to stabilize throughout the topology over time, and the combined ``wisdom'' of the cluster should be increasingly prioritized over the deleterious effect of local learning. As such, the method of Delta Sum Learning is summarized as:

\begin{equation}
    w_{a,t_0 + T} =  \frac{ \sum^{N+1}_{n=0} w_{n,t_0}}{N+1} + \lambda(t_0 + T) \sum^{N+1}_{n=0}\Delta w_{n,t_0+T} 
\end{equation}
\begin{equation}
\label{eq:interpol}
     \lambda(t) =  min\left(A + \frac{t}{B}, C\right)
\end{equation}

Where A, B and C are constants to be empirically determined for optimal training. While B is strongly dependent on the expected maximum number of training epochs up to which the factor should increase, which may not be calculable during continuous or online training, functional values for both A and C are determined for the scenario in the evaluation section.

%

\section{Architecture}
\label{ch:flocky}

To enable gossip cluster building and to evaluate Delta Sum Learning in a realistic scenario, the model is implemented in the architecture illustrated by Fig. \ref{fig:arch}. This architecture uses Flocky\cite{Goethals2025} as a basic framework as its services allow for the discovery of other nodes, as well as providing a decentralized cluster-wide metadata store and orchestration (Swirly and deployment) services. 

Flocky is an open source framework that leverages Open Application Model (OAM) to model both node capabilities and workload requirements. Each node monitors its own resources, capabilities, and running workloads (i.e. Traits and Components, respectively) through Capability Providers (e.g. Workload agent and ML Service in Fig. \ref{fig:arch}), which are stored in the Repository Service and shared with nodes \circled{1} found by the Discovery Service. Not all nodes in a cluster are mutually known; the Discovery Service is configured to explore nodes only within certain parameters (e.g. latency, physical distance, ...), so each node only directly knows a small part of the whole cluster. Workloads are requested through the Swirly Service, which matches requirements with known node capabilities from the Repository Service, and subsequently deploys the workloads on the most suitable node through its Deployment Service, which in turn leverages a workload agent (e.g. Docker\footnote{Docker Containers - https://www.docker.com/} or Feather\cite{goethals2024feather}) to run them.

To enable Gossip Learning within Flocky, this research adds two additional services to manage model updates through gossiping:

The \textbf{ML Service} monitors running ML workloads for model updates consisting of a base model and a delta tensor \circled{2}. Currently, this is implemented through shared memory to improve performance, although a REST endpoint may be less error prone at the cost of possibly significant overhead (i.e. model serialization). Additionally, this service acts as a Capability Provider, mapping the ML capabilities of the local node and reporting them to the Repository Service. When a model update is detected, it is reported to the Gossip Service along with the workload identifier for further dissemination in the cluster. Conversely, whenever the Gossip Service reports a model update for a specific workload, it is retrieved and passed along with the source node identifier to the workload for integration. This service is currently implemented in Golang; for performance reasons, integration methods are implemented in the workloads themselves. 

The \textbf{Gossip Service} leverages a Golang implementation of the SWIM gossip protocol to enable gossiping \circled{3}. Whereas the Discovery and Repository services constitute a gossip-like cluster, these are only used for OAM-based node and workload data, while the Gossip Service may be used for generic and operational workload data. Furthermore, using SWIM guarantees the propagation of workload data to relevant nodes, whereas the base Flocky services are designed to be error tolerant and ``best effort''. Unlike normal gossip clusters, the Gossip Service cluster is not created through configuration or joining; the nodes discovered by the Discovery Service are periodically synchronized as comprising the gossip cluster. The Gossip Service is generic: when a message is received (i.e. model update from ML Service), it forwards the message using its key (i.e. workload identifier) to other nodes in the cluster. Listeners can subscribe to the Gossip Service by providing a list of keys, e.g. the ML Service providing a list of its monitored workloads.

The use of these services is enabled by adding a specific Trait to the Flocky framework, allowing workloads to be designated as Gossip Learning ML workloads. This Trait requires an endpoint within the workload (shared memory file or REST service) to exchange model updates, and when the Deployment Service applies this Trait it notifies the ML service of a new workload to be monitored.

\section{Evaluation}
\label{ch:evaluation}

This section describes the evaluation setup and any implementation details that deviate from Section \ref{ch:flocky}, as well as the evaluation scenario and methodology.
The code for all services and custom evaluation tools is made available on GitHub\footnote{https://github.com/idlab-discover/flocky}.

\subsection{Evaluation Setup}

Evaluations are performed on a single Ubuntu 24.04 desktop PC with an Intel Core Ultra 7 155H processor, 32GiB dual channel LPDDR5, and a 1TB M.2 NVMe drive.

\subsection{Evaluation Scenario}

The effectiveness of Delta Sum Learning is measured by comparing it to standard model averaging and variance corrected \cite{tian2024vanishing} averaging. Although many other integration methods exist, Delta Sum Learning can theoretically enhance every method that can be modified for separate base model and update dissemination. For performance reasons, all alternatives are implemented in the Python ML workload itself. 
Semi-random topologies are generated containing 10, 25 and 50 nodes, inspected to avoid bisection and to ensure that each node has between 1 and 8 neighbours for direct communication (Discovery and gossip). Average connectivity for the topologies is determined to be 3.3, 3.2 and 4.2 neighbours, respectively.
Each node runs Discovery, Repository, and Gossip/ML services, and an ML workload. The ML workload consists of a Convolutional Neural Network (CNN) comprised of two iterations of convolution and max pooling, followed by two dense layers, and uses the default MNIST digits classification dataset\cite{MNIST} for training. The model contains 55,338 parameters, for a total size of 221,352 bytes at fp32.
The ML workload runs in training mode for 200 epochs, delivering model updates to the ML service, with integrations every 20 epochs. After training mode, it continues in convergence mode until round 235 to gauge the effect of additional integration rounds. During convergence rounds, full models are averaged rather than using Delta Sum Learning (or other strategies), as there is no longer any model delta.

\subsection{Methodology}
\label{ch:methodology}

All nodes in an evaluation scenario are simulated on the same machine by assigning incremental service ports, allowing discovery at localhost rather than distinct IP addresses.
Training evaluation metrics are exported by the ML workload after each training epoch, and for each gossip update round after epoch 200. While absolute simultaneity is not guaranteed by the evaluation setup, the results show that equal CPU prioritization and timing individual processes using sleep operations is enough to ensure limited divergence in epoch count at any time across all simulated nodes. Each simulated node is assigned an equal share of the MNIST dataset, divided into a training portion and validation portion, as well as the full validation dataset to gauge global convergence.
The dynamic gossip learning factor $\lambda$ constructed in Eq. \ref{eq:interpol} is used during training; for the MNIST scenario, the constants $A = 0.15$, $B = 1000$, and $C = 0.35$ are used. Both lower (A = 0.05 and C = 0.1) and higher (A = 0.35 and C = 0.6) learning factors result in worse performance for the specified scenario, although metaparameter optimization may provide slightly different values.
Memory use was determined to be significantly higher when using hardware acceleration ($\sim$1.5GiB) than when running models on CPU only ($\sim$900MiB). Although some memory is shared across processes, the ML workloads are run entirely on CPU to ensure the feasibility of 50 node topologies. No artificial delays are used other than a short sleep operation intended to synchronize epochs with rounds; other than that, training epochs are executed as fast as possible, making evaluation strongly CPU bound. As such, CPU use is not monitored, while measured network traffic in larger topologies is not reliable due to significant throttling of gossip traffic resulting from CPU starvation. As a result, gossip rounds tend to lag training epochs in larger topologies.
For all evaluations, CPU scaling is disabled to ensure a proper comparison. For practical purposes and to avoid timing issues, the ML workloads are run as host processes rather than containers. As a result, the Workload agent is stubbed, while the rest of the components (e.g. Discovery and Repository) function normally.

\section{Results}
\label{ch:results}

\begin{figure*}[htbp]
\centering
\pgfplotstableread[col sep=comma,]{data/f10-e1metrics_average_proc.csv}{\ftenaveraging}
\pgfplotstableread[col sep=comma,]{data/f10-e1metrics_varcorr_proc.csv}{\ftenvar}
\pgfplotstableread[col sep=comma,]{data/f10-e1metrics_deltacorr_proc.csv}{\ftendelta}
\pgfplotstableread[col sep=comma,]{data/f25-e1metrics_averaging_proc.csv}{\ftfaveraging}
\pgfplotstableread[col sep=comma,]{data/f25-e1metrics_varcorr_proc.csv}{\ftfvar}
\pgfplotstableread[col sep=comma,]{data/f25-e1metrics_deltacorr_proc.csv}{\ftfdelta}
\pgfplotstableread[col sep=comma,]{data/f50-e1metrics_averaging_proc.csv}{\ffifaveraging}
\pgfplotstableread[col sep=comma,]{data/f50-e1metrics_varcorr_proc.csv}{\ffifvar}
\pgfplotstableread[col sep=comma,]{data/f50-e1metrics_deltacorr_proc.csv}{\ffifdelta}
\begin{tikzpicture}
\pgfplotsset{set layers}
\begin{groupplot}[
  group style={
    	group size=3 by 1,
	ylabels at=edge left,
    	group name=plots,
    	horizontal sep=1.5cm,
	vertical sep=1.8cm
  },
	xlabel={Epoch/round},
	ylabel={Accuracy},
]
\nextgroupplot[	
	width=0.35\linewidth,
    	legend style={at={(1.6,-0.4)},
	  anchor=north,legend columns=3},
	xmin=0,
	xmax=235,
	ymin=0.95,
 	ymax=0.992,
	enlarge x limits=0.01,
	title={a) 10 node topology}
]
\addplot [red!100] table [x=index, y=test_acc_median]{\ftenaveraging};\addlegendentry{Averaging}
\addplot[blue!100] table [x=index, y=test_acc_median]{\ftenvar};\addlegendentry{Variance corr.}
\addplot [green!40!black] table [x=index, y=test_acc_median]{\ftendelta};\addlegendentry{Delta sum}

\addplot [red!60,name path=lower, fill=none, draw=none] table [
    x=index, y=test_acc_min]{\ftenaveraging};
\addplot [red!60,name path=upper, fill=none, draw=none] table [
    x=index, y=test_acc_max]{\ftenaveraging};
\addplot[red!60,fill opacity=0.5] fill between[of=lower and upper];

\addplot [blue!60,name path=lower, fill=none, draw=none] table [
    x=index, y=test_acc_min]{\ftenvar};
\addplot [blue!60,name path=upper, fill=none, draw=none] table [
    x=index, y=test_acc_max]{\ftenvar};
\addplot[blue!60,fill opacity=0.5] fill between[of=lower and upper];

\addplot [green!60,name path=lower, fill=none, draw=none] table [
    x=index, y=test_acc_min]{\ftendelta};
\addplot [green!60,name path=upper, fill=none, draw=none] table [
    x=index, y=test_acc_max]{\ftendelta};
\addplot[green!60,fill opacity=0.5] fill between[of=lower and upper];

\nextgroupplot[
	width=0.35\linewidth,
	xshift=-0.5cm,
	xmin=0,
	xmax=235,
	ymin=0.95,
 	ymax=0.992,
	enlarge x limits=0.01,
	title={b) 25 node topology}
]

\addplot [red!100] table [x=index, y=test_acc_median]{\ftfaveraging};
\addplot[blue!100] table [x=index, y=test_acc_median]{\ftfvar};
\addplot [green!40!black] table [x=index, y=test_acc_median]{\ftfdelta};

\addplot [red!60,name path=lower, fill=none, draw=none] table [
    x=index, y=test_acc_min]{\ftfaveraging};
\addplot [red!60,name path=upper, fill=none, draw=none] table [
    x=index, y=test_acc_max]{\ftfaveraging};
\addplot[red!60,fill opacity=0.5] fill between[of=lower and upper];

\addplot [blue!60,name path=lower, fill=none, draw=none] table [
    x=index, y=test_acc_min]{\ftfvar};
\addplot [blue!60,name path=upper, fill=none, draw=none] table [
    x=index, y=test_acc_max]{\ftfvar};
\addplot[blue!60,fill opacity=0.5] fill between[of=lower and upper];

\addplot [green!60,name path=lower, fill=none, draw=none] table [
    x=index, y=test_acc_min]{\ftfdelta};
\addplot [green!60,name path=upper, fill=none, draw=none] table [
    x=index, y=test_acc_max]{\ftfdelta};
\addplot[green!60,fill opacity=0.5] fill between[of=lower and upper];

\nextgroupplot[
	width=0.35\linewidth,
	xshift=-1cm,
	xmin=0,
	xmax=235,
	ymin=0.95,
 	ymax=0.992,
	enlarge x limits=0.01,
	title={c) 50 node topology}
]

\addplot [red!100] table [x=index, y=test_acc_median]{\ffifaveraging};
\addplot[blue!100] table [x=index, y=test_acc_median]{\ffifvar};
\addplot [green!40!black] table [x=index, y=test_acc_median]{\ffifdelta};

\addplot [red!60,name path=lower, fill=none, draw=none] table [
    x=index, y=test_acc_min]{\ffifaveraging};
\addplot [red!60,name path=upper, fill=none, draw=none] table [
    x=index, y=test_acc_max]{\ffifaveraging};
\addplot[red!60,fill opacity=0.5] fill between[of=lower and upper];

\addplot [blue!60,name path=lower, fill=none, draw=none] table [
    x=index, y=test_acc_min]{\ffifvar};
\addplot [blue!60,name path=upper, fill=none, draw=none] table [
    x=index, y=test_acc_max]{\ffifvar};
\addplot[blue!60,fill opacity=0.5] fill between[of=lower and upper];

\addplot [green!60,name path=lower, fill=none, draw=none] table [
    x=index, y=test_acc_min]{\ffifdelta};
\addplot [green!60,name path=upper, fill=none, draw=none] table [
    x=index, y=test_acc_max]{\ffifdelta};
\addplot[green!60,fill opacity=0.5] fill between[of=lower and upper];
\end{groupplot}
\end{tikzpicture}
\caption{ Accuracy for Delta Sum Learning and alternative strategies for topology sizes ranging from 10 (a)) to 50 (c)). Shaded areas indicate the full range of accuracies of all nodes, except statistical outliers for visualization purposes. Comparing the different topology sizes, Delta Sum Learning results in a far lower accuracy loss as gossip topology size increases than alternative strategies. }
\label{fig:accuracy}
\end{figure*}
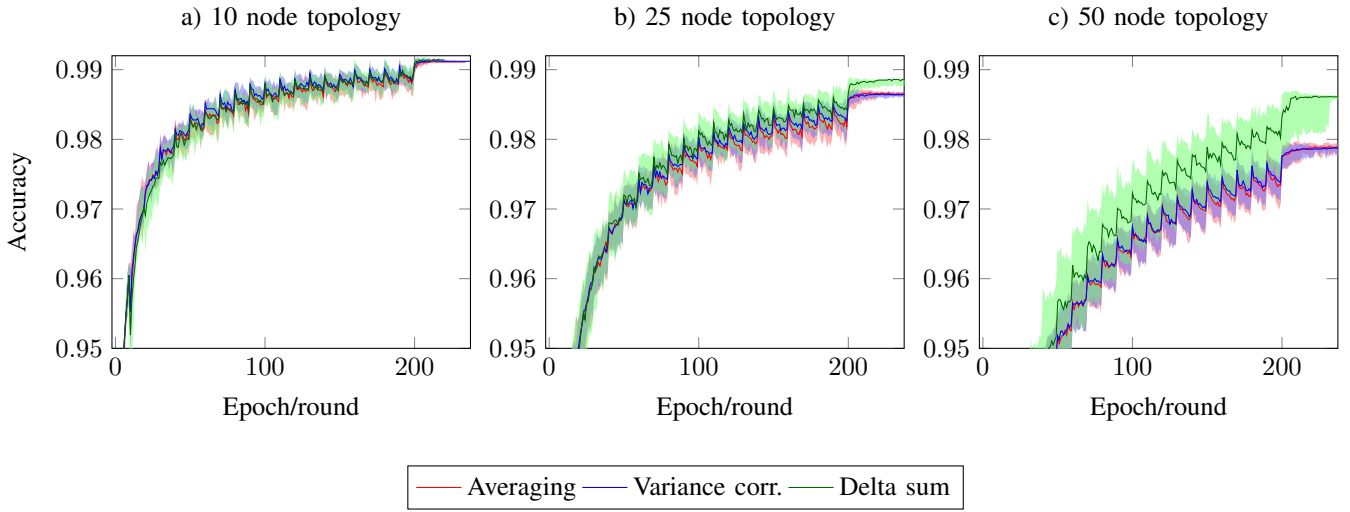

\begin{figure}[htbp]
\begin{tikzpicture}
\pgfplotsset{set layers}
\begin{axis}[
    enlarge x limits=0.05,
    ylabel near ticks,
	ylabel={Accuracy},
	xlabel={Number of nodes},
	y tick label style={
	        /pgf/number format/.cd,
	        precision=3
	    },
	xtick=data,
    	ymin=0.97,
    	legend style={at={(0.5,-0.2)},
	  anchor=north,legend columns=3},
]
\addplot+[red,mark options={fill=red},error bars/.cd,
y dir=both,y explicit]
coordinates {
(10,0.991100013) -= (0,0) += (0,0)
(25,0.986500024795532) -= (0,0.000300049781799983) += (0,0.000199973583220991)
(50,0.978900015354156) -= (0,0.00139999435415605) += (0,0.000599980354308971)
};
\addplot+[blue,mark options={fill=blue},error bars/.cd,
y dir=both,y explicit]
coordinates {
(10,0.9911999702453612) -= (0,0) += (0,0)
(25,0.986400008201599) -= (0,0.00040000677108798) += (0,0.000499963760375977)
(50,0.978699982) -= (0,0.000299989999999917) += (0,0.000500024000000043)
};
\addplot+[green!40!black,mark options={fill=green!40!black},error bars/.cd,
y dir=both,y explicit]
coordinates {
(10,0.9914000034332277) -= (0,0) += (0,0)
(25,0.988499999046325) -= (0,0.000699996948241965) += (0,0.000100016593932994)
(50,0.986149996519088) -= (0,0.000950008630752008) += (0,0.000149995088578048)
};
\legend{Averaging,Variance corr.,Delta sum}
\end{axis}
\end{tikzpicture}
\caption{ Median accuracy of Delta Sum Learning compared to other strategies at round 235, for an increasing number of nodes in the topology. While accuracy loss for other strategies is almost linear with number of nodes, Delta Sum appears to follow a logarithmic pattern. }
\label{fig:stratcompare}
\end{figure}
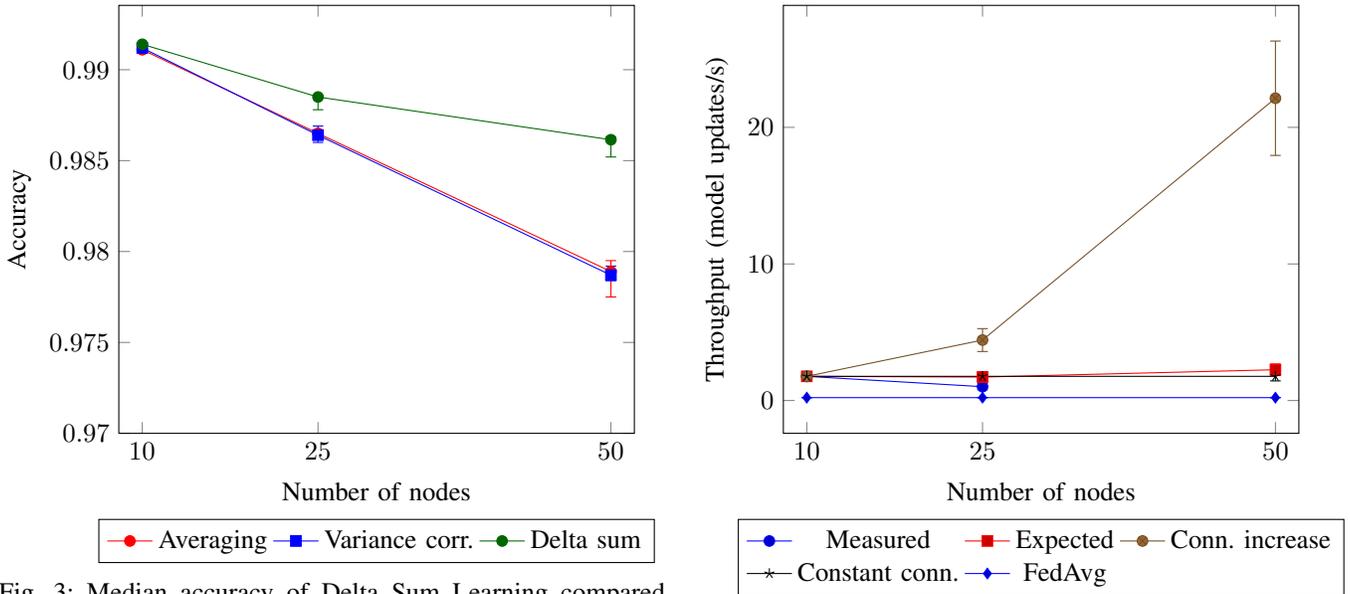

\begin{figure}[htbp]
\begin{tikzpicture}
\pgfplotsset{set layers}
\begin{axis}[
    enlarge x limits=0.05,
    ylabel near ticks,
	ylabel={Throughput (model updates/s)},
	xlabel={Number of nodes},
	xtick={10,25,50},
    	legend style={at={(0.5,-0.2)},
	  anchor=north,legend columns=3},
]
\addplot+[error bars/.cd,
y dir=both,y explicit]
coordinates {
(10,1.77065882205598) -= (0,0.335026985057811) += (0,0.335026985057811)
(25,1.01336006032444) -= (0,0.22280278732464) += (0,0.22280278732464)

};
\addplot+[error bars/.cd,
y dir=both,y explicit]
coordinates {
(10,1.77065882205598) -= (0,0.335026985057811) += (0,0.335026985057811)
(25,1.71700249411489) -= (0,0.324874652177271) += (0,0.324874652177271)
(50,2.25356577352579) -= (0,0.426397980982669) += (0,0.426397980982669)

};
\addplot+[error bars/.cd,
y dir=both,y explicit]
coordinates {
(10,1.77065882205598) -= (0,0.335026985057811) += (0,0.335026985057811)
(25,4.42664705513994) -= (0,0.837567462644528) += (0,0.837567462644528)
(50,22.1332352756997) -= (0,4.18783731322264) += (0,4.18783731322264)

};
\addplot+[error bars/.cd,
y dir=both,y explicit]
coordinates {
(10,1.77065882205598) -= (0,0.335026985057811) += (0,0.335026985057811)
(25,1.77065882205598) -= (0,0.335026985057811) += (0,0.335026985057811)
(50,1.77065882205598) -= (0,0.335026985057811) += (0,0.335026985057811)

};
\addplot+[error bars/.cd,
y dir=both,y explicit]
coordinates {
(10,0.21) -= (0,0) += (0,0)
(25,0.21) -= (0,0) += (0,0)
(50,0.21) -= (0,0) += (0,0)

};
\legend{Measured,Expected,Conn. increase,Constant conn.,FedAvg}
\end{axis}
\end{tikzpicture}
\caption{ Network throughput in model updates per second as measured, and calculated for various theoretical scenarios such as Constant connectivity, Connectivity increase due to node density, and FedAvg under the same conditions as the evaluations. }
\label{fig:neteff}
\end{figure}

This section presents the results of the evaluations, considering both training metrics and networking aspects. Although network throughput was not explicitly measured due to issues discussed in Section \ref{ch:methodology}, theoretical performance can be compared to FL. For training, only the accuracy metric is represented. Although other metrics such as loss and F1 scores may provide a better indication for most models, loss was observed to be completely (inversely) in line with accuracy\footnote{https://github.com/idlab-discover/flocky/evaldata}, and accuracy is sufficient to gauge a limited model.

\subsection{Accuracy and Convergence}

Fig. \ref{fig:accuracy} shows the accuracy of Delta Sum Learning over the entire runtime of the evaluation scenarios compared to standard averaging and variance corrected averaging. For small 10-node topologies, the difference between all strategies is neglegible; variance corrected averaging has a higher performance during the early stages but all models end up with identical accuracy a few integration rounds after the last training epoch. Importantly, accuracy ends up at 99.1\%, providing a baseline for larger topologies. 
At 25 nodes, Delta Sum Learning results in a higher accuracy overall. Integrations every 20 epochs have a higher positive impact, while the negative effect of local training is less pronounced due to the dynamic gossip learning factor. The latter advantage is shared by variance corrected averaging, although standard averaging and variance corrected averaging result in equal accuracy at round 235. 
The 50 node topology illustrates the advantages of Delta Sum Learning in terms of scaling and global convergence. Despite each node having a minute share of the full training dataset, at epoch 50 Delta Training already results in a significantly higher accuracy than the alternatives, with integrations having a higher impact than in the 25 node topology. Additionally, local training has a significantly lower negative impact on global validation dataset accuracy. Furthermore, with Delta Sum Learning it only takes a few convergence rounds after training for median accuracy in the topology to improve to the top performing nodes, while the worst performing nodes catch up by round 230. While other strategies also converge to some degree, they retain a significant range above and below median accuracy beyond the end of the evaluation.

Fig. \ref{fig:stratcompare} compares the results at round 235 across topology sizes. Both standard averaging and variance corrected averaging sustain a linear accuracy loss as topologies grow, with variance corrected averaging showing a more uniform accuracy across larger topologies. Delta Sum Learning, however, has the same accuracy as the alternative strategies in a 10 node topology, but loses accuracy logarithmically as topology size grows: absolute accuracy drops from 99.1\% at 10 nodes to 98.6\% at 50 nodes, while the alternatives end up at 97.9\%. Importantly, the results also show that the effect of the dynamic learning factor from Eq. \ref{eq:interpol} is independent of node connectivity; while the 25 node topology has a lower average connectivity and the the 50 node topology a higher connectivity than the 10 node topology, Delta Sum Learning performs increasingly better than alternatives with a growing number of nodes.

\subsection{Network Efficiency}
\label{ch:scalability}

Various expected trends for network efficiency are shown in Fig. \ref{fig:neteff}, expressed in model updates per second as a model-agnostic unit. Due to the discovery algorithm, network traffic is expected to scale as $O(r^2 \delta_n)$, with discovery distance r and local node density $\delta_n$, although the gossip protocol may be configured to limit the number of updates sent per round. Calculated trends are based on the Measured values for the 10 node topology; as illustrated by the chart the 25 node topology is significantly CPU limited, resulting in a real-time throughput drop as CPU intensive training takes longer to complete. Due to the manual checks imposed on the evaluation topologies, connectivity is relatively consistent across all scenarios, resulting in the Expected throughput adhering closely to expected throughput for Constant connectivity (i.e. either a physically growing topology, or $r^2=1/\delta_n$ if density increases). The trend for Connectivity increase indicates throughput for increasing node density in a topology of constant physical size and $r$. Finally, the expected network load for FedAvg is shown based on  a model update every 5 seconds, and a model synchronization every 20 updates, in line with the 10 node GL scenario. 

\section{Future Work} 
\label{ch:discussion}

The results show that Delta Sum Learning results in sigifnicantly higher accuracy and better global convergence than standard averaging or variance corrected averaging applied to Gossip Learning. Furthermore, Delta Sum Learning exhibits a far lower accuracy loss than alternatives as topologies grow, despite limited connectivity and model updates. However, some aspects of the proposed solution may be further improved. 

The Delta Sum Learning model examines the impact of model integration from first hop (i.e. direct neighbour) updates. However, it does not take into account training data asymmetry and model divergence beyond direct neighbours; instead, it uses a gossip learning factor to mitigate these issues. While this factor is also necessary to regulate the balance between local learning and remote updates, further examination of the impact of connectivity, model divergence during training, and data asymmetry may improve the model.

GL avoids bottlenecks by removing centralized integration, technically allowing scaling to larger training topologies than FL. Conversely, the evaluation shows GL requires significantly more network traffic than FL in all (theorized) scenarios. While decentralized systems that require global consistency necessarily have a significant communication overhead, the absolute overhead may be reduced by, for example, modifying the gossip protocol to limit the number of updates sent per training epoch depending on connectivity, or limiting updates based on when model integrations are scheduled to occur. Furthermore, the Flocky discovery mechanism can be configured for lower connectivity, and the gossip protocol itself may be amenable to improvements for the specific use case of GL.

Finally, the gossip service may be improved to create a separate cluster per model, or per namespace (i.e. group of models or workload metadata). This approach avoids synchronization of specific data with nodes that have no use for it, which improves network efficiency and reduces security risks by restricting data to only those nodes that require it. In terms of security the implemented method of Delta Learning borrows some concepts from Secure Aggregation, as updates obfuscate any training data and, as with reducing network overhead, could be averaged over several epochs instead of distributing each epoch update to further reduce low-level information leakage.

\section{Conclusion}
\label{ch:finito}

This paper introduces Delta Sum Learning to improve global convergence and training accuracy. It provides a mathematical model to characterize the operation and benefits of Delta Sum Learning, and presents a practical implementation which operates in Flocky, a decentralized orchestration framework. The evaluation section compares the implementation to standard averaging and variance-corrected averaging GL approaches using a basic CNN trained on the MNIST digits classification dataset. Evalution results indicate that Delta Sum Learning performance is increasingly better than the alternative approaches as topologies grow, resulting in a logarithmic loss of accuracy rather than a linear loss, as well as providing global convergence close to the best performing nodes instead of converging to median cases. Concretely, the global accuracy loss for Delta Sum Learning is 58\% lower than the alternatives when scaling from 10 node to 50 node topologies. However, results also show that GL requires around 5 times more traffic between nodes to achieve global synchronization than FL, which scales with node connectivity but may be reduced through future work.
Finally, some topics for future work are discussed concerning the accuracy, network overhead and security aspects of Delta Sum Learning and its implementation in Flocky.







\bibliographystyle{IEEEtran}
\bibliography{IEEEabrv,biblio}



\end{document}